# A Game-Theoretic Model of Demand Response Aggregator Competition for Selling Stored Energy in Regulated and Unregulated Power Markets

Mahdi Motalleb, Reza Ghorbani[1]

**Abstract**

This work is concerned with the application of game-theoretic principles to model competition between demand response aggregators for selling excess energy stored in electrochemical storage devices directly to other aggregators in a power market. This market framework is presented as an alternative to the traditional vertically-integrated market structure, which may be better suited for developing demand response and smart-grid technologies, in addition to increasing penetration of independent renewable energy generation devices. Demand for power generated by the utility through combustion of fuel could be replaced, lowering emission of pollutants, when the energy used to charge the batteries is produced sustainably and traded on smaller scales. The four variants of game are considered: both non-cooperative (unregulated competition) and Stackelberg (regulations on transaction price and size), each with and without DR scheduling. The Nash equilibrium is derived for each game variant in order to serve as a bid-price decision making criteria which determines the optimal bidding strategy for an aggregator to sell in the market. The model is applied to a case study involving completion for selling between two aggregators. Bidding strategy is dependent on parameters inherent to an aggregator's energy storage hardware, and the strategy selected by each aggregator does not vary with the variations in the game conditions considered. Demand response scheduling offers greater payoff for aggregators who implement it, compared with those who do not. Addition of transaction price and volume regulations to the market do not affect the participants optimal bidding strategies (the Nash equilibrium), but lowers payoffs for all aggregators participating in the market relative to unregulated competition.

## 1. Introduction

Development and implementation of smart grid technologies offers substantial advantages over traditional vertically integrated electric utilities. Both parties stand to benefit from proliferation of these technologies. The end-users of electrical power primarily wish to minimize the cost they pay to the utility for the energy they require to power their devices. Opposed to this, the utility company wishes to maximize its net profit and is concerned with load scheduling as they must remain able to provide adequate supply. In this paper, a market framework has been developed based on game-theoretical concepts where Demand Response Aggregators (DRAs) compete with each other to sell energy stored in consumer's battery systems. Our model determines the optimal bidding decision for each DRA to maximize its own pay off under the conditions of an incomplete information game.

Game theory has been used in power system markets to interpret a participant's behavior in deregulated environments and to allocate costs among pool participants. Two different hybrid algorithms were presented [1, 2, 3] for the Generation Expansion Planning (GEP) problem for a pool-based electric market where the modified-game-theoretic algorithms were divided into two programing levels: master and slave. A static computational game theoretic model has also been developed [4] to investigate the impacts of competition on the wholesale price of electricity, the demand for electricity, the profits of firms, and levels of various polluting emissions. A Medium Run Electricity Market Simulator (MREMS) based on game theory was presented in [5]. This simulator incorporated two different games, one for the unit commitment of thermal units and one for strategic bidding and hourly market clearing. The most common electricity bidding mechanisms in electricity auction markets were analyzed using signaling game theory [6] and also a Swarm platform was used to develop a simulation model based for multiple agents. The role of sustainable energy volatility was investigated in the context of market participant's competitive expansion planning problem in [7]. An incomplete information non-cooperative game-theoretic method where each generation company (GENCO) perceives strategies of other market participants was applied to make decisions about strategic generation capacity expansion. An agent-based game-theoretic formulation was presented in [8] to investigate the learning speed of traders and their strategic collaboration in dynamic electricity markets. It was shown that the learning speed of traders decreases during large fluctuations in the power exchange market.

[1] Mahdi Motalleb and Reza Ghorbani, are with Renewable Energy Design Laboratory (REDLab) at the Department of Mechanical Engineering, University of Hawai'i Manoa, Honolulu, Hawai'i, 96822. Phone: +1-808-956-2292, Fax: +1808-956-0767. E-mail: motalleb@hawaii.edu, rezag@hawaii.edu.

The Authors of [9] expanded the application of an Equilibrium Problem with Equilibrium Constraints (EPEC) model, typically used with multi-leader-follower games, to a spatial market. In that model, the prevalent market setting was investigated in the international market for metallurgical coal between 2008 and 2010, whose market characteristics provide arguments for a wide variety of alternate market structures. A two-sided agent-based framework was proposed in [10] employing a game-theoretic/Particle Swarm Optimization (PSO) hybrid simulation approach to provide a suitable platform for sustainable GEP. A Bayesian game model analyzing cross-border transfer of electric power was presented in [11] to explain cross-border transfer within a market containing players controlling unreliable marginal generators. A dynamic game-theoretic model was developed in [12] to analyze the impacts of natural gas market reformation in promoting Natural Gas-Fired Electricity (NGFE) generation, in which hourly Real-Time Pricing (RTP) was applied in both natural gas and electricity markets [53]. An innovative game theoretic framework was proposed in [13] for a next-generation retail electricity market ("Energy Internet") with high penetration of distributed residential electricity suppliers ("Energy Cells"). The envisioned Energy Internet proposes a large number of distributed Renewable energy generation and energy storage systems connected to the grid through plug-and-play interfacing devices. A game-theoretic framework for economic operations of future residential distribution systems was presented in [14] for cases involving extensive participation of distributed electricity prosumers which defined novel roles for both the utilities and "customers". The game theoretic algorithms used to determine the retail electricity market clearing price consider group coalition scenarios of multiple electricity prosumers.

Game theory is a useful mathematical tool to handle problems related to demand side management (DSM) [15, 48-51]. Several game-theoretical demand response (DR) programs have been proposed with differing objectives such as: determining the optimal hourly incentive to be offered to customers who sign up for load curtailment [16, 17, 52], scheduling load usage by creating several possible tariffs for consumers such that net demand remains below some threshold [18], adjusting demand to meet supply, as well as smoothing the aggregated load in the system [19], and evaluating the impact of the response capability of smart-home consumers on promoting further distributed PV penetration [20]. Game theory was used in [21] to present an optimization model to minimize load curtailment needed to restore equilibrium to the operating point when the system is in a fault condition (e.g., loss of generation). An optimal time-of-use pricing with an evolutionary game-theoretic perspective was proposed in [22] for urban gas markets where a power structure demand response program was employed to simulate user demand response.

Decision-making (DM) processes are the principle product of game theory. Basically, Game theory is the formal study of DM under competitive conditions where choices potentially affect the interests of the other players. In [23] a DM approach for Liquefied Natural Gas (LNG) projects was presented. The approach was based on a consensus algorithm addressing the consensus output over a common value using cost functions within a framework based on game theory. In another work [24], a DM process for conceptual planning and project evaluation in the oil and gas industry was presented where the set of strategic decisions is generated by a binary genetic algorithm. In [25], DM processes of industrial and environmental concerns was evaluated with a game theoretic approach where Industry and Environment were considered as two players with conflicting interest to find optimal strategies in governing energy policy. A bi-level, complete-information, matrix game-theoretic model was proposed in [26] to assess the economic impact and make operational decisions in carbon- constrained restructured electricity markets. A game theoretic modeling approach was performed in [27] to develop financial transmission rights bidding strategies for power suppliers assuming that they have adequately forecast Locational Marginal Prices (LMPs). The game theoretic model considered multiple participants as well as network contingencies. An evolutionary imperfect-information game approach was proposed in [28] to analyze bidding strategies in electricity markets with price-elastic demand. The research work presented in [29] characterized the impact of Long-term plans on short-term maintenance decisions of GENCOs by applying the Cournot model, which has been used for strategic generation dispatch of generating units in electricity markets.

In order to characterize the DM process, Nash equilibrium is used. A set of strategies is a Nash equilibrium if no player can do better by unilaterally changing his or her strategy [30]. The authors of [31] suggested to use a theory of evolutionary games and the concept of a "near Nash equilibrium" to simulate the electricity market in the presence of more than two producers. In [32], a novel game-theoretic approach was presented to the Generation Maintenance Scheduling (GMS) problem in electricity markets where a coordination procedure for an Independent Service Operator (ISO) was modeled and the GENCOs GMS process was modeled as a non-cooperative dynamic game, and the GENCO's optimal strategy profile was determined by the Nash equilibrium of the game. In [33], the conventional model of a dynamic Cournot electricity market game (where GENCOs are assumed to hold a uniform and accurate belief concerning market dynamics) was replaced by a more realistic model with subjective demand errors coining a new equilibrium concept termed Subjective Equilibrium (SE). System performance, equilibrium output, profit and customer surplus, at SE were analyzed. The results of [33] suggested that the system equilibriums are strongly influenced by the GENCOs' knowledge about market demand.

In game theory, different types of games are utilized for analysis of different types of problems. The different types of games are categorized by number of players involved, symmetry of the game, and whether or not, cooperation among players is allowed. In the literature for power markets, the different game models used include: cooperative [34, 35, 36, 37], non-cooperative [20, 38], Stackelberg [9, 39], multi-leader-follower [9], Forchheimer (one leader) [5], and Bertrand games (all players are leaders) [5]. Beside the application in power markets, game theory has been used in diverse, and related fields such as: analysis of Electric Vehicle (EV) charging station construction [40], charging method for plugged in hybrid EVs [41], analysis of power grid vulnerability [42], performance evaluation of thermal power plants [43], analysis of effects of higher domestic gas prices in Russia on the European gas market [44], and interactive energy management of networked microgrid-based active distribution systems [45].

In the present work, a demand response market framework has been developed based on game-theoretic considerations. In this market, the commodity is energy stored in consumers' batteries. End users of electricity charge their batteries when the electricity is cheap and sell it back to the grid (or other consumers) when the electricity is more expensive. There are four basic concepts in game theory including: players, nodes, moves, and payoffs. In the proposed market framework, the players are DRAs and each node belongs to one DRA. The possible moves of the game are the player's decision–i.e.: any of the options chosen by a given DRA. A player's strategy will determine the action the player will take at each stage of the game. In the proposed market structure, strategies are different biddings of DRAs. The payoff to each player (DRA) is the difference between monetary gain upon selling its stored energy and the cost of storing that energy. A comprehensive quadratic cost function has been proposed in this study for discharging the energy stored in batteries. Also, it has been assumed that players are privy to no information about other player's moves. Therefore, our problem is cast as a real, incomplete information game ($i$-game). Two different types of game are considered here: non-cooperative, and Stackelberg. In the non-cooperative game, DRAs compete to sell the energies stored in the batteries without any limitation or constraints in transaction size or price; but in the Stackelberg game, the utility (as game leader) controls the DR market through restricting both. In each game, two different DR programs are considered: non-scheduled demand and price-sensitive scheduling of water heaters (or all thermostatic devices). Dynamic Programing (DP) has been used for price-sensitive scheduling.

Section 2 explains some principles of game theory's application to energy markets. In section 3, a cost function and a payoff function have been developed in order to utilize in the proposed method. Section 4 describes the proposed game-theoretic market framework including the alternate game types and DR schedules considered. The presented market framework has been applied to a case study and its dissection and results have been provided in section 5. Finally, conclusions and possible future works are presented in section 6.

## 2. Principles and definitions of game theory in power market

First, let us briefly present the fundamental principles and definitions of the game-theoretic concepts we need to analyze a power market. The following subsections cover: a proposed function we will use to model the aggregator's cost, the bidding process of DRAs, and the payoff function for DRAs.

### 2.1. Game theory in power market

In game theory, beliefs are formulated against risky alternatives in order to maximize the expected revenue (payoff function) for aggregators. In the real competition, each player (DRA) remains unaware of the detailed data of other players (for instance data of strategies and payoffs). Probability theory gives an expected value of the payoff given the probability of events and Bay's law is used to revise beliefs given new information. The objective of the game-theoretic analysis is to find the Nash equilibrium – the best strategy from the set of possibilities in the sense that no incentive exist to alter the equilibrium strategy despite the strategies chosen by the other players. Fig. 1 illustrates a power market where some of the aggregators acting as buyers and others acting as sellers submit their bids to a center (ISO) and this center sets a market spot price between aggregators.

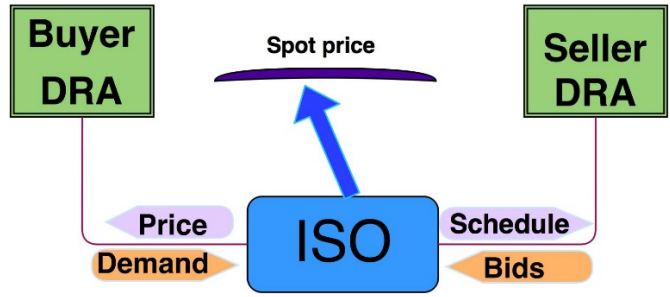

Fig. 1: General structure of a power market including buyer DRAs, seller DRAs, and ISO

DRAs participate in the market to sell their available stored energy which is not consumed in powering local loads. In a power transaction game, aggregators' transactions are modeled as a game of strategies to maximize their payoffs. Two types of games are considered in this research: non-cooperative and Stackelberg. A non-cooperative game takes place when each player (DRA) is interested to maximize its own payoff and there is no cooperation between aggregators to coordinate their strategies. In a non-cooperative game, some aggregators choose a strategy while other aggregators try to identify their best response to that strategy. In a Stackelberg game, there is one leader (utility) to limit the market price and power transferred to ensure all policies are respected in the market.

### 2.2. Aggregator cost function

This subsection shows how DRA's cost function can be presented with a quadratic function. Since the only energy resources available for sale by a given DRA is the the aggregation of stored energy in residential batteries, the aggregator cost function is the battery cost function (cost of discharging the battery). For the battery, the function we chose is a variant of the widely used logarithmic barrier function, used as a penalty function in interior point methods [46]:

$$C_{h_i}^{bat}\left(\Delta E_{h_i}\right) = -a_{h_i}\log(1-\frac{\Delta E_{h_i}}{B}) \qquad (1)$$

where $C_{h_i}^{bat}$ is cost of the battery in house $i$ as a function of stored energy, $a_{h_i}$ is a pricing coefficient determined by the utility to give higher prices during peak-hours, $\Delta E_{h_i}$ is the total load (kWh), and $B$ is a parameter that has been used to give cost values very close to the values given by a quadratic one, it also serves as the maximum typical value for $\left|\Delta E_{h_i}\right|$. The relation between the presented cost function (Eq. 1) and the quadratic one can be understood from its Taylor expansion. Since $\Delta E_{h_i}/B < 1$, the Taylor series expansion is:

$$C_i^{bat}\left(\Delta E_{h_i}\right) = -a_{h_i}\log(1-\frac{\Delta E_{h_i}}{B}) \approx -a_{h_i}(-\frac{\Delta E_{h_i}}{B} - \frac{\Delta E_{h_i}^2}{B^2}) = \frac{a_{h_i}\Delta E_i}{B} + \frac{a_{h_i}\Delta E_i^2}{B^2} \quad (2)$$

Updates in load data resolve every 15 minutes from measurement devices. We assume that the load is constant for each 15 minutes between updates and the quadratic cost function of the battery can be rewritten as a function of power ($\Delta P_{h_i}$):

$$C_{h_i}^{bat}\left(\Delta P_{h_i}\right) = \frac{a_{h_i}(0.25\Delta P_{h_i})}{B} + \frac{a_{h_i}(0.25\Delta P_{h_i})^2}{B^2} \quad (3)$$

In order to obtain the coefficients $a_{h_i}$ and $B$, two factors should be considered: the electricity price of the grid (for charging the battery) and the capital and maintenance costs of the battery. Thus, the cost of the battery of house $i$ for charging the power by an amount $\Delta P_{h_i}$ is:

$$C_{h_i}^{bat}\left(\Delta P_{h_i}\right) = C_{h_i}^{grid}\left(\Delta P_{h_i}\right) + C_{h_i}^{C\&M}\left(\Delta P_{h_i}\right) \quad (4)$$

where $C_{h_i}^{grid}\left(\Delta P_{h_i}\right)$ is the cost which house $i$ pays to the utility to sell its electricity and to charge its own battery as amount power of $\Delta P_{h_i}$ and $C_{h_i}^{C\&M}\left(\Delta P_{h_i}\right)$ is the contribution of capital and maintenance costs of the battery of house $i$ for power of $\Delta P_{h_i}$. The coefficients $a_{h_i}$ and $B$ can be obtained using Eq. 4. Since the individual house cost functions are quadratic in form, aggregation of the houses will also produce quadratic cost functions for each DRA:

$$C_g(P_g) = v_g P_g + u_g P_g^2 \quad (5)$$

where $C_g$ is the cost function of aggregator $g$ for selling the stored power of $P_g$ (in $kW$), and the coefficients of $u_g$ and $v_g$ are:

$$v_g = \frac{0.25 a_g}{B} \quad (6)$$

$$u_g = \frac{a_g(0.25)^2}{B^2} \quad (7)$$

where $a_g = \frac{1}{N_h}\sum_{i=1}^{N_h} a_{h_i}$, and $N_h$ is the number of houses under aggregator $g$.

### 2.3. Bidding process of DRAs

This section explains how DRAs bid their price for selling the stored energy in batteries, based on the cost function defined in section 2.2. From Eq. 5, the marginal cost of aggregator $i$ is:

$$\lambda_i = \frac{dC_i}{dP_i} = v_i + 2u_i P_i \qquad (8)$$

Since marginal cost is a linear function of stored power in batteries, we assume that aggregators' bids are also linear functions of stored power:

$$\lambda_i = \lambda_{0i} + m_i P_i \qquad (9)$$

where $\lambda_i$ Is the bid price per unit of power at discharging power level $P_i$, $\lambda_{0i}$ is the marginal cost of electricity at $P_{0i}$, and $m_i$ is the slope of the bid curve. The market coordinator (ISO) receives sale bids from the DRAs given by Eq. 9 and matches the lowest bid with the buyer aggregators. In Fig. 2, the aggregator intends to increase the battery discharging beyond $P_{0i}$ if the spot market price is greater than $\lambda_{0i}$. Here, $T_i$ is the net power interchanged; if $T_i$ is positive (negative), the aggregator is selling power to (buying power from) the other aggregators, $-P_{0i} \leq T_i \leq P_{max,i} - P_{0i}$. The maximum stored power in the aggregated storages of DRA $i$ is $P_{max,i}$, the minimum of the local load of DRA $i$ which should be fed by local batteries is $P_{0i}$ (can be established as one of the governing policies of aggregator $i$).

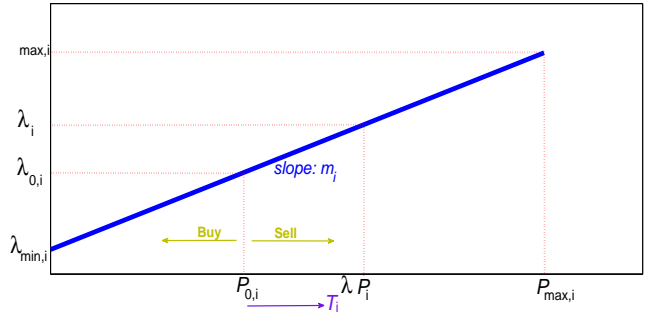

Fig. 2: Aggregator $g$'s biding curve

### 2.4. Aggregator's payoff function

The mathematical formulation for each DRA's payoff function in the proposed game-theoretic framework is described below. For a given spot market price, $\phi$, the *ith* DRAs payoff, $R_i$, is:

$$R_i = -\Delta C_i + \phi T_i \qquad (10)$$

where $\Delta C_i$ is the difference between the discharging cost of the batteries in aggregator before and after power transactions, and $T_i$ is the net power transacted.

Let $T$ denote a power transaction; associated with each $T$ is a pair of aggregator payoffs $(\delta, \omega)$ for participants 1 and 2. The Nash bargaining problem for two aggregators is denoted by $[\Re, (\delta, \omega)]$ characterized by a region, $\Re$, and a payoff point $(\delta, \omega)$ in $\Re$. Without any trading, the payoff is $(0,0)$; a trade would take place if and only if both aggregators agree upon a solution in $\Re$ which represents an acceptable deal for the participants involved. The problem is formulated as follows: given $(\delta_0, \omega_0) = (0,0)$ in $\Re$ as the initial solution, find a different solution, $(\delta_1, \omega_1)$, that satisfies the conditions:

- $(\delta_1, \omega_1)$ is a point of $\Re$.
- $\delta_1 \omega_1 \geq \delta_0 \omega_0 \quad \forall (\delta_0, \omega_0) \in \Re \ \& \ \delta_1 > 0, \omega_1 > 0$

The point $(\delta_1, \omega_1)$ is the Nash equilibrium solution to the bargaining game $[\Re, (\delta, \omega)]$.

From the definition of the payoff function for DRA $g$ in eq. 10, the objective function is:

$$\text{maximize} \sum_{k \in K} \prod_{p_{ij}, T_{ij}} R_{ij}^k (p_{ij}, T_{ij}),$$

where

$$R_{ij}^k (p_{ij}, T_{ij}) = \phi_{ij} T_{ij}^* - [C_i(P_{gi} + T_{ij}) - C_i(P_{gi})] \quad (11)$$

$R_{ij}^k$ is the payoff of DRA $i$ (as a seller) after transmitting the power of $T_{ij}$ to aggregator $j$ (and receiving the power of $T_{ij}^*$ by aggregator $j$) in contract $k$. The difference between $T_{ij}$ and $T_{ij}^*$ is due to transmission losses. With negligible transmission losses $T_{ij} = T_{ij}^*$. The parameter, $\phi_{ij}$, is the transaction price and $P_{gi}$ is the amount of the local loads of the aggregator $i$ which is fed by the local storages before power transaction. The first term in eq. 11 is the gross revenue of the aggregator due to the transaction and the second bracketed term is the change in the aggregator's batteries discharging cost owing to the transaction. In the case of two aggregators shown in Fig.3 (aggregator 1 as seller and 2 as buyer), the objective function is:

maximize $\{R_{seller} \times R_{buyer}\}$,

where

$$R_{seller} = \phi_T T_{12}^* - [C_1(P_{g1} + T_{12}) - C_1(P_{g1})] \quad (12)$$

$$R_{buyer} = -\phi_T T_{12}^* - [C_2(P_{g2} - T_{12}^*) - C_2(P_{g2})] \quad (13)$$

and $\phi_T$ is the tranction price and $R_{seller}$ and $R_{buyer}$ are the aggregators payoffs. Each aggregator has a constrained storage capacity, and payoffs are constrained by $R_{seller} \geq 0$, and $R_{buyer} \geq 0$, indicating that negative payoffs are excluded. When there are no transactions, the payoff for each aggregator is zero.

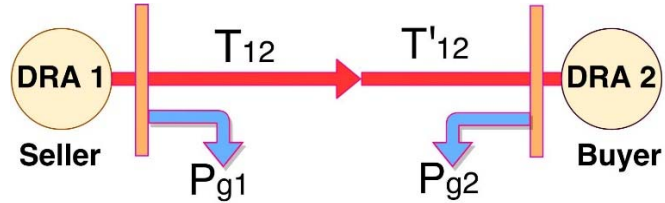

Fig. 3: A system including two DRAs as seller and buyer

In real competition between aggregators, participants don't have information about the other participants such as cost function coefficients, availability of the batteries, and biding prices. The following section shows the proposed game-theoretic market framework for competition between DRAs to sell energy stored in batteries in an incomplete information game.

## 3. Proposed game-theoretic market framework for DRAs

The proposed methodology is explained in this section. The first subsection includes details about the incomplete information game to make a market competition among DRAs. The proposed market framework based on game theory has been presented in the second subsection. All mathematical and probabilistic details for the developed market framework (including two different game types each with and two demand scheduling programs) have been provided in second subsection.

Fig. 4 shows a visual schematic of the proposed market framework. Each DRA manages a number of buildings with control signals sent through a wireless network. Each house contains various devices possibly including Water Heaters (WHs), Batteries, Air Conditioners (ACs), and Electric Vehicles (EVs). The DRAs decide how to bid for buying and selling power in the market. A DP scheduler outputs signals for deciding when a WH should run-on/off setting as a function of time-to satisfy the conditions in the game (social welfare in all games, and price-sensitive DR WH scheduling in the games where it is implemented) given as inputs hot water usage data (in all games) and possibly electricity price signals from the utility (only for the games with price-sensitive DR WH scheduling). The DRAs communicate through a wireless network with servers which communicate with the market directly to arrange transactions. In the Stackelberg games, the servers also mitigate size and price of transactions in accordance with utility policy constraints which are transmitted over the wireless network.

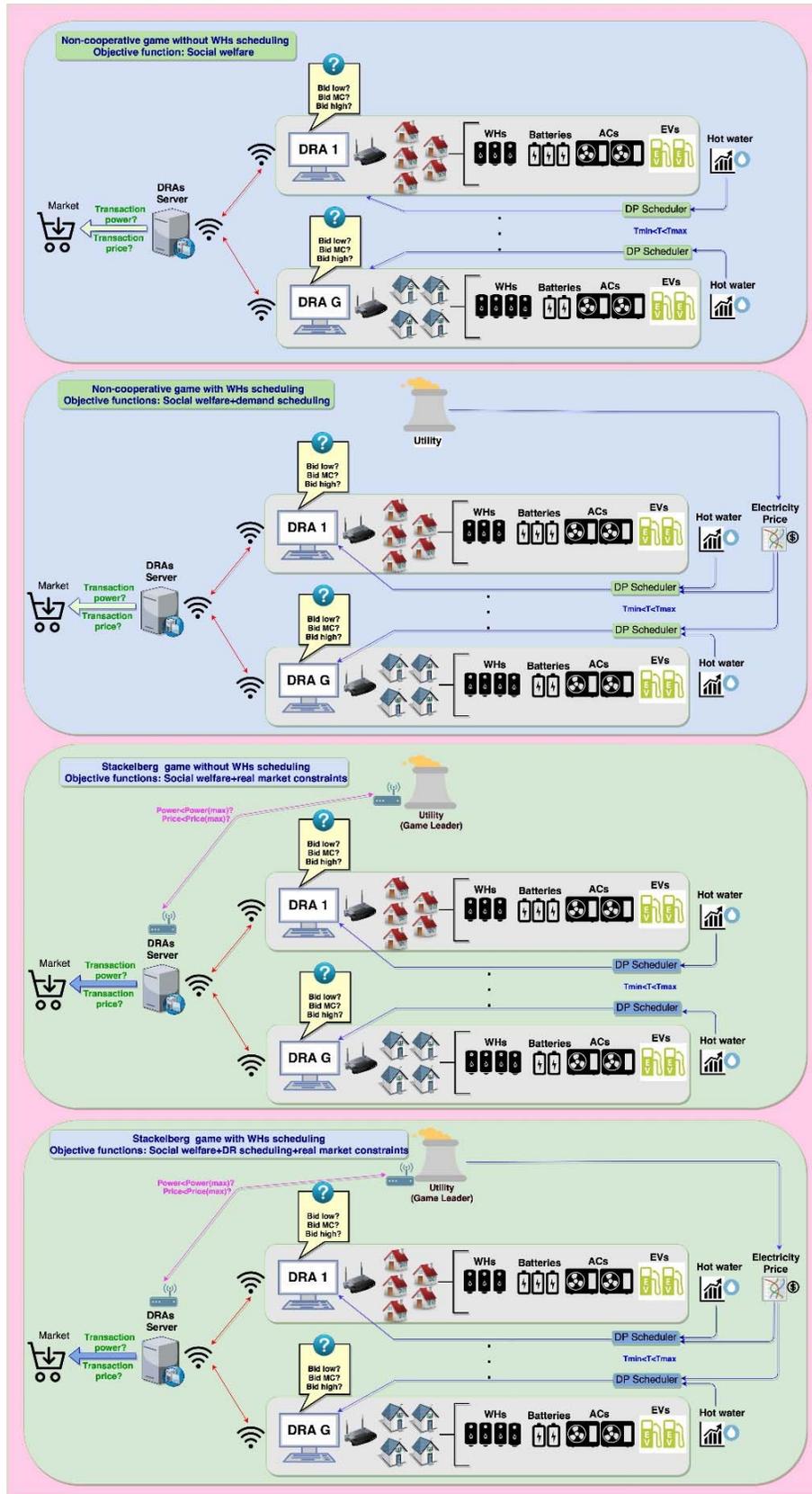

Fig. 4: General schematic of the proposed market framework in two types of games and two DR scheduling program

### 3.1. Market competition between DRAs with incomplete information

In this section, a market model is proposed to model competition among aggregators in an electricity market where participants have incomplete information. Based on eq. 10, each aggregator's payoff is a function of storages discharging costs, power transactions, and the spot market price. Hence, an individual aggregator's payoff is a function of bids offered by other aggregators. Each participant estimates the other participants' bids in order to maximize its own payoff. Each aggregator has complete information concerning its own payoff but lacks information critical to predicting others actions precisely, this competition is known as an incomplete game (*i*-game). Here, an aggregator's unknown characteristics are modeled by classification of participant's types. The DRA's type contains information concerning: its own payoff function, other aggregators' pay off functions, grid electricity prices, availability of charged batteries, etc.. An aggregator's type corresponds to its battery discharging cost structure, that is, coefficients $a_i$ and $B$ (Eq. 5 and 6). Each aggregator would have full knowledge of its own costs, but only an estimate of the remaining aggregator' costs. Each participant adjusts the slope of *m* in its bidding curve (Fig. 2) in order to maximize its payoff, $R_i$, in Eq. 10. The choice of *m* corresponds with the bidding strategy (bid low, bid high, bid marginal cost) in the game. The main goal of the game is to determine the optimal DM process used to determine what value of *m* to choose in order to maximize payoff despite the unknown parameters of the other participants.

The DRAs use a Bayesian approach to deal with incomplete information. In this approach, a probability distribution, $\Pi$, represents the unknown parameters. The expected value of the payoffs of the aggregators are maximized for $\Pi$. In estimating the probability distribution, each aggregator uses only the information common to all aggregators.

Consider a scenario described from DRA *B*'s perspective for competing with DRA *A* for selling the stored energy of the batteries to aggregator *C* depicted in Fig. 5, below. We assume that the participants' types are drawn at random from hypothetical populations $\Theta_A$ and $\Theta_B$ containing types $t_A^m$ and $t_B^n$, respectively where $m = 1,...,M$ and $n = 1,...,N$ index the possible types of DRA *A* and *B*, respectively. For instance, to model the uncertainty in *A*'s discharging cost, *B* would assume that there are *M* possible types of *A*, $t_A^m$, each with its corresponding cost. Generally, aggregator *B* knows *n* but does not know *m*, and hence does not know its opponent's type (i.e. it does not know *A*'s discharging costs). Each DRA estimates the probability distributions, $\Pi$, for the random variable of the "type" of its competitors based on freely available published information such as electricity prices, demand curves, water usage (from water heaters), availability of storages, and aggregators' parameters.

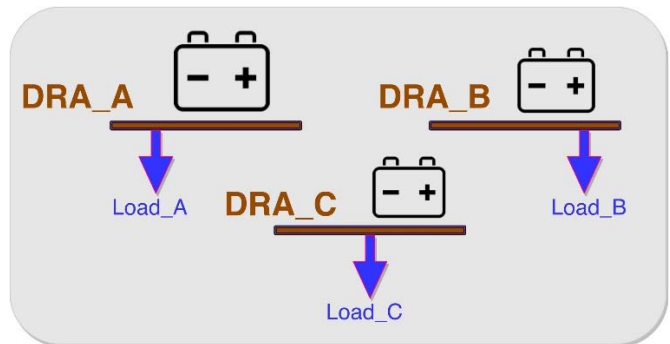

Fig. 5: A grid including three DRAs: *A* and *B* as sellers and *C* as buyer

### 3.2. Game-theoretic market framework
For our purposes, the goal of the game is to calculate the expected payoffs for different strategies for aggregators *A* and *B* (in the mentioned scenario) to find the Nash equilibrium and to choose the strategy that maximizes payoff:

$$EP_A^m = \sum_{n=1}^{N} \eta_A^m(n).H_A^m(s_A^m, s_B^n, m, n) \tag{14}$$

where $EP_A^m$ is the expected payoff function of aggregator $A$ if it is of type $m \in M$. $n$ is the type of opponent aggregator $B$ ($n \in N$). The conditional probability $\eta_A^m(n)$ is the probability DRA A is of type $m$ for DRA B of type $n$:

$$\eta_A^m(n) = prob(t_B^n | t_A^m) = \frac{\pi_{mn}}{\sum_{n=1}^{N} \pi_{mn}} \tag{15}$$

where $prob$ shows the conditional probability function, and $\pi_{mn}$ is the basic probability distribution $\prod$ corresponds to the probability that $A$ is type $m$ and $B$ is type $n$. A DRA's strategy (for bidding) is determined by its type. $s_A^m$ is a vector of strategies for $A$'s type $m$. For example, a vector of strategies may be: bid high, bid at marginal cost, and bid low. $H_A^m$ is the conditional payoff of DRA $A$ which depends on the strategies of DRA A, $s_A^m$, and DRA B, $s_B^n$. Aggregator $A$ needs to know the opponent's type to maximize $H_A^m$, and since the information is not available in an *i-game*, aggregator $A$ maximizes the expected value of the payoff $EP_A^m$.

Similarly, for aggregator $B$, the expected payoff function is:

$$EP_B^n = \sum_{m=1}^{M} \eta_B^n(m).H_B^n(s_B^n, s_A^m, n, m) \tag{16}$$

where

$$\eta_B^n(m) = prob(t_A^m | t_B^n) = \frac{\pi_{mn}}{\sum_{m=1}^{M} \pi_{mn}} \tag{17}$$

### 3.2.1. Non-cooperative game

The *i-game* considered previous converts to a $(M + N)$-participant game as an imperfect information complete game *(c-game)*, where $M$ and $N$ are the number of different types of aggregators $A$ and $B$, respectively, in the case that each aggregator knows the others' payoff functions and, basic probability distributions involved in the game but does not know the opponent's type. The Nash equilibrium is the solution. We consider two variants of (*c*-game): non-cooperative and Stackelberg, in a scenario involving three DRAs competing to sell energy stored in their batteries. In each game, the effect of price-sensitive demand response scheduling (for WHs and other similar thermostatic storages) has been considered and the market model has been proposed with/without DR scheduling. The algorithm shown (Fig. 6) depicts the procedure used to calculate the requisite parameters used in obtaining the Nash equilibrium.

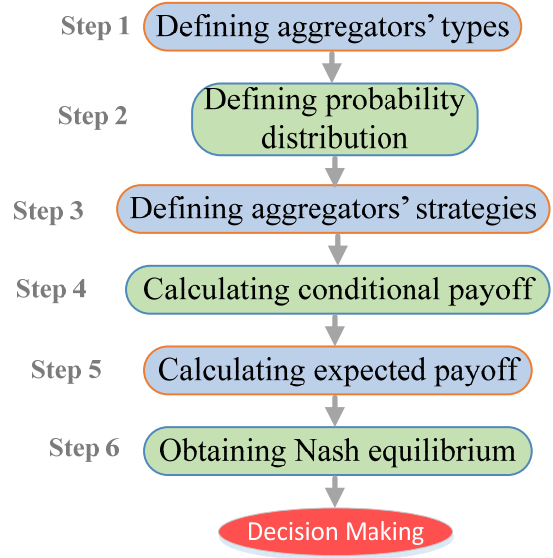

Fig. 6: steps of calculating required parameters in the game-theoretic market

The following subsections explain how to determine the parameters needed in the DM process for each game and each demand response scheduling program.

### 3.2.1.1. Non-cooperative game without DR scheduling

Assuming that DRAs A, and B, know C's payoff function, we may determine the nash equilibrium following the procedure outlined in Fig. 6.

*Step1: Define aggregators' types:*

In the first step, a set of different possible discharging cost function's coefficients ($a_i$ and $B$ in Eq. 5 to 7) is defined for each aggregator $A$ and $B$. The number of defined types are $M$ and $N$ for aggregators $A$ and $B$, respectively. The participant's type is determined largely by electricity price and availability of the batteries for discharging. The electricity price is significant because the batteries are charged directly from the grid. Availability of stored energy in the batteries may be estimated by considering use patterns for high power consumption devices such as Water Heaters (WHs) and other thermostatic devices. For example, when the WHs are consuming power (based on the water usage and water temperature data), most of the energy stored in the batteries is being consumed locally to power these devices. Considering these two factors we may classify an aggregator into one of four scenarios:

$\sigma_1: \{elec_{\exp}, WH_{off}\}$

$\sigma_2: \{elec_{\exp}, WH_{on}\}$

$\sigma_3: \{elec_{ch}, WH_{off}\}$

$\sigma_4: \{elec_{ch}, WH_{on}\}$

For a non-cooperative game without price-sensitive demand response, scheduling depends entirely on water usage and is completely independent of the price of electricity since the thermostat functions only to maintain the social welfare demands of the customer without considering price. When two events are independent, the probability of both events occurring is the product of the probabilities of each event. For aggregator scenarios $\sigma_1$, $\sigma_2$, $\sigma_3$, $\sigma_4$, we have:

$$P(\sigma_1) = P(elec_{exp} \cap WH_{off}) = P(elec_{exp}) * P(WH_{off})$$
$$P(\sigma_2) = P(elec_{exp} \cap WH_{on}) = P(elec_{exp}) * P(WH_{on})$$
$$P(\sigma_3) = P(elec_{ch} \cap WH_{off}) = P(elec_{ch}) * P(WH_{off})$$
$$P(\sigma_4) = P(elec_{ch} \cap WH_{on}) = P(elec_{ch}) * P(WH_{on})$$

$\sum_{f=1}^{4} P(\sigma_f) = 1$, and $P(\sigma)$ is the probability of event $\sigma$. The probability of the WHs' power state ($P(WH_{off})$ and $P(WH_{on})$) are obtained from the DP scheduler block described in Fig. 4. In this case, the scheduler is determined by typical water consumption profile and the only objective function is to maintain social welfare- i.e. that the WH maintains temperatures in the range require by the customer to ensure sufficient hot water is available.

Let $\Psi_A^f$ and $\Psi_B^f$ be probability distributions modeling uncertainties in each aggregator's discharging cost:

$$\Psi_A^f = [\Psi_A^f(1), \Psi_A^f(2),..., \Psi_A^f(M)] \qquad (18)$$

$$\Psi_B^f = [\Psi_B^f(1), \Psi_B^f(2),..., \Psi_B^f(N)] \qquad (19)$$

$\Psi_A^f(m)$ is the probability that aggregator $A$ is of type $m$ ($m \in M$) in scenario $f$ and $\Psi_B^f(n)$ is the probability that aggregator $B$ is in type $n$ ($n \in N$).

*Step2: Define the basic probability distribution of the game:*

The probability that $(m, n)$ would represent participants $A$ and $B$'s types, respectively, would depend on electricity price and local demand (WHs' conditions). The expected probability that $A$ is type $m$ and $B$ is type $n$, $\pi_{mn}$, is defined as:

$$\pi_{mn} = \sum_{f=1}^{4} (P(\sigma_f).\vartheta(\sigma_f).\Psi_A^f(m).\Psi_B^f(n)) \qquad (20)$$

Where $\vartheta(\sigma_f)$ is a participation coefficient for scenario $f$: equal to one for participants selling energy in the market and 0 otherwise. For example, $\vartheta(\sigma_2) = 0$ because when the electricity is expensive and the water usage is high, energy stored in electrochemical cells will be used preferably for powering local loads. $\vartheta(\sigma_1) = \vartheta(\sigma_3) = \vartheta(\sigma_4) = 1$ because if the price of electricity is cheap, the stored energy is available for sale as loads may be supplied directly from the grid at lower cost; whereas even if the price is high, absence of local demand leaves energy available for sale. Given this scheme for predicting $\pi_{mn}$, the original *i*-game may be treated as a *c*-game with imperfect information. Each aggregator now knows its cost function's coefficients and may compute its own discharging costs without knowing the opponent's discharging cost. In this step $\eta_A^m(n)$ and $\eta_B^n(m)$ are calculated ($n \in N, m \in M$).

*Step 3: Define aggregators' strategies*

An aggregator's type corresponds to a set of strategies defined by bid slopes (Fig. 2). We assume that each aggregator has three strategies based on Eq. 9. The slope of the bid curve of $g^{th}$ aggregator (Fig. 2) is $m_g = \varepsilon_S \times u_g$ where $\varepsilon_S$ is set to $[\varepsilon_{S1}, \varepsilon_{S2}, \varepsilon_{S3}]$ for three different strategies. These strategies are: $\varepsilon_{S1} < 2$ for bidding less than marginal cost, $\varepsilon_{S2} = 2$ for biding at marginal cost, and $\varepsilon_{S3} > 2$ for bidding more than marginal cost. Coefficient $u_g$ was defined in Eq. 5. In this step $s_A^m$ and $s_B^n$ are calculated ($n \in N, m \in M$).

*Step 4: Define aggregators' conditional payoff*

In this step the following system of equations has been solved for a combination of participant types and strategies:

$$\begin{cases} \lambda_A = \lambda_B = \phi_T \\ P_A + P_B = L_C \end{cases} \quad (21)$$

where $P_A$ and $P_B$ are the power remaining in the storages cells of DRA A and B, after feeding their local loads. This power is sold to aggregator C ($L_C$ is the power purchased by aggregator C), and $\phi_T$ is the transaction price. After determining $P_A$, $P_B$, and $\phi_T$, in eq. 21, the conditional payoffs of DRA $A$ and $B$ may be calculated with eq. 12 and 13. The conditional payoff matrixes take the form:

$$H_A^m(n) = \begin{bmatrix} h_{A,11} & h_{A,12} & h_{A,13} \\ h_{A,21} & h_{A,22} & h_{A,23} \\ h_{A,31} & h_{A,32} & h_{A,33} \end{bmatrix} \quad (22)$$

$$H_B^n(m) = \begin{bmatrix} h_{B,11} & h_{B,12} & h_{B,13} \\ h_{B,21} & h_{B,22} & h_{B,23} \\ h_{B,31} & h_{B,32} & h_{B,33} \end{bmatrix} \quad (23)$$

In $H_A^m(n)$, each row corresponds to a strategy of $A$ for type $m$ and each column corresponds to a strategy of $B$ for type $n$. For instance, $h_{A,23}$ in $H_A^2(1)$ corresponds to $A$'s type 2 payoff if $A$ decides to bid marginal cost against a situation where $B$ is type 1 and bids above marginal. $H_B^n(m)$ is defined similarly.

*Step 5: Define expected payoff matrices.*

Expected values of the payoffs, $EP_A^m$ and $EP_B^n$, are calculated using eq. 14, 15, 16, 22 and 23. The final form of the $EP_A^m$ matrix is:

$$EP_A^m = \begin{bmatrix} \kappa_{11}^1 & \kappa_{12}^1 & \kappa_{13}^1 & \kappa_{21}^1 & \kappa_{22}^1 & \kappa_{23}^1 & \kappa_{31}^1 & \kappa_{32}^1 & \kappa_{33}^1 \\ \kappa_{11}^2 & \kappa_{12}^2 & \kappa_{13}^2 & \kappa_{21}^2 & \kappa_{22}^2 & \kappa_{23}^2 & \kappa_{31}^2 & \kappa_{32}^2 & \kappa_{33}^2 \\ \kappa_{11}^3 & \kappa_{12}^3 & \kappa_{13}^3 & \kappa_{21}^3 & \kappa_{22}^3 & \kappa_{23}^3 & \kappa_{31}^3 & \kappa_{32}^3 & \kappa_{33}^3 \end{bmatrix}$$

Each column of $EP_A^m$ corresponds to the presumed strategy of participant $A$'s opponent. For instance, $\kappa_{32}$ column in $EP_A^1$ is $A$'s type 1 payoff when $B$ uses strategy 3 against $A$'s type 1 and strategy 2 against $A$'s type 2. The same notation applies to $EP_B^n$.

*Step 6: Obtain the Nash equilibrium of strategies*

Utilizing $EP_A^m$ and $EP_B^n$, the Nash equilibrium pairs are obtained. We look for the collection of strategies in which each aggregator's strategy would be represented by the best response to other aggregator's strategies. The Nash equilibrium is a prediction of how the game will be played. An aggregator's optimal bid is derived for this equilibrium point. All aggregators predict that a particular Nash equilibrium will occur and there is no incentive to play differently.

### 3.2.1.2. Non-cooperative game with DR scheduling

In this game, price-sensitive scheduling of water heaters is considered. The steps of this game are the same as non-cooperative game without DR scheduling (section 3.2.1.1) with a single exception in the first step. For a non-cooperative game without DR scheduling, WHs keep the water temperature in a given range to satisfy social welfare constraints, and situation of WHs (On/Off) is independent of the electricity price. In non-cooperative game with DR scheduling, the objectives are both social welfare and demand scheduling. The first step in finding the Nash equilibrium of the non-cooperative game with DR scheduling requires knowledge of conditional probabilities since the events are not independent.

*Step1: Define aggregators' types*

As in section 3.2.1.1, a set of possible discharging cost function's coefficients ($a_i$ and $B$ in eq. 6 and 7) is defined for each DRA acting as a seller (A and B). Similarly we define four scenarios ($\sigma_1,...,\sigma_4$) depending on the same variables two variables (electricity price and WHs' condition), but in this game these events are dependent due to price-sensitive DR scheduling. When two events are dependent, the probability of both occurring is defined using conditional probability:

$$P(\sigma_1) = P(elec_{exp} \cap WH_{off}) = P(elec_{exp}) * P(WH_{off} / elec_{exp})$$

$$P(\sigma_1) = P(elec_{exp} \cap WH_{on}) = P(elec_{exp}) * P(WH_{on} / elec_{exp})$$

$$P(\sigma_3) = P(elec_{ch} \cap WH_{off}) = P(elec_{ch}) * P(WH_{off} / elec_{ch})$$

$$P(\sigma_4) = P(elec_{ch} \cap WH_{on}) = P(elec_{ch}) * P(WH_{on} / elec_{ch})$$

The second part of each equation is the conditional probability. For instance $P(WH_{off} / elec_{exp})$ is the probability that the water heater is off given that electricity is expensive. These conditional probabilities are calculated using the DP scheduler (mentioned in Fig. 4) considering two objective functions: social welfare and DR scheduling. In this game more batteries are available to provide power because scheduling has disabled some loads, thus each DRA stands to make a larger payoff then in the game without DR scheduling (3.2.1.1).

### 3.2.1.3. Stackelberg game without DR scheduling

In the previous sections while considering non-cooperative games, we assumed that the transaction price (and transaction power) depend only on the DRAs bid prices (eq. 21). DRAs with storage capabilities will tend to charge their batteries when energy is less expensive during off-peak hours, in order to sell it to other DRAs during peak hours when the price is high thereby maximizing their gains per transaction. When enough DRAs do this, a phenomenon termed "Reverse Peaks" becomes apparent where the shift of peak consumption is accompanied by a high tendency of aggregators to sell their stored energy back during high price hours. This undesirable effect can be resolved by allowing a market controlling center (i.e., utility or any other center controlled by the utility) to a play the role of market controller in the game. The leader is allowed to set bounds on market transaction quantity, and price thereby ameliorating the undesirable effects on the market associated with reverse peaks. Competition including a market controller can be modeled as a Stackelberg game. A Nash equilibrium exist for the Stackelberg game-i.e. each rational player will chose a particular predictable strategy which offers them no incentive to alter it despite the strategies chosen by the competitors.

Finding the Nash equilibrium for the Stackelberg game follows the procedure outlined in fig. 6, similar to the previous sections with the exception that in calculating the conditional payoff (step 4), the transaction powers ($P_A, P_B$) and transaction price ($\phi_T$) are controlled by the leader (utility). If these values, ($P_{A_{max}}, P_{B_{max}}, \phi_{T_{max}}$), fall within the range permitted by the controller, the transactions proceeds; but if the obtained values from eq. 21 are higher than maximum allowable values, these maximum values are used to clear the market. Applying the utility's policies in Stackelberg game, prevents the huge effects of players (aggregators) in the market and also prevents reverse peaks.

### 3.2.1.4. Stackleberg game with DR scheduling:

The analysis of the Stakleberg game with DR scheduling follows reasoning similar to one without DR scheduling (section 3.2.1.3) with the exception that we must assume that electricity price and water usage are dependent events. Conditional probabilities are used in calculating the the probabilities of the intersection of dependent events for the possible aggregator scenarios in step 1 analogous to what was done in section 3.2.1.2 when considering DR scheduling for a non-cooperative game.

## 4. Case study, Discussion and results

For illustration purposes, consider a system with three DRAs labeled *A, B*, and *C* as depicted in fig. 7. Load data for these three DRAs was taken from a real grid model on the island of Maui (Hawaii-United States) [47].

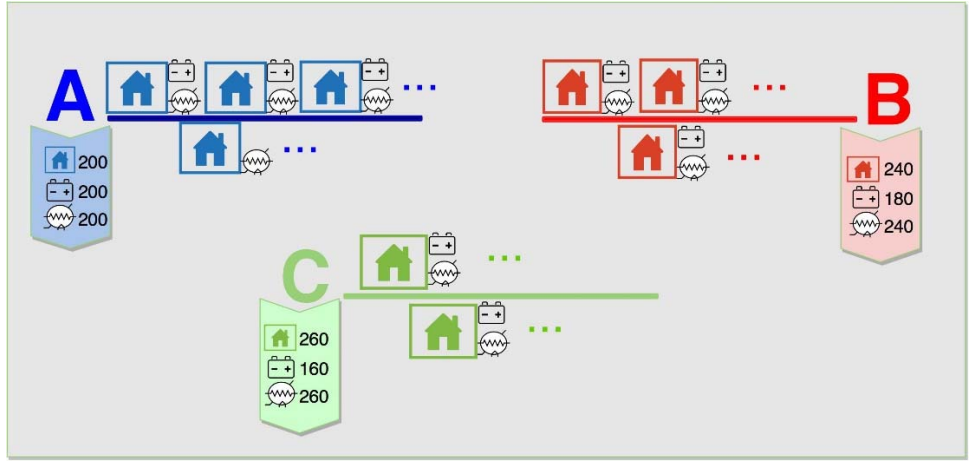

Fig. 7: Case study including 3 agregators: *A* and *B* as sellers and *C* as buyer

DRA A manages 200 houses, all of which contain electrochemical storage cells. DRA B manages 240 house, but only 180 have storage devices. DRA C manages 240 houses of which only 180 are equipped with storage devices. All houses have WHs assumed to draw the majority of the load (60%). The nominal power of each residential storage device is 3.3 kW. Thus, the maximum generation capacity of any DRA is:

$$P_A^{gen} = 660 kW, P_B^{gen} = 594 kW, P_C^{gen} = 528 kW$$

In this case study we assume that the nominal power of each WH is 4.5 kW. Thus, the total power demand of *A* is:

$$P_A^{demand} = 200*4.5*10/6 + P_A^{gen} = 2160 kW$$

Similarly, for other aggregators:

$$P_B^{demand} = 2394 kW, P_C^{demand} = 2478 kW$$

Since the marginal cost of DRA *C* is greater than for other aggregators, A and B will compete to sell power to C. We assume that $M = 2$ and $N = 2$ and that the DRAs discharge cost coefficients are: Seller_A $\begin{Bmatrix} a_A = [4, 4.3] \\ B = [25, 23.5] \end{Bmatrix}$

Seller_B $\begin{Bmatrix} a_B = [4.2, 4.5] \\ B = [24, 23.5] \end{Bmatrix}$ Buyer_C $\begin{Bmatrix} a_C = 5 \\ B = 21.5 \end{Bmatrix}$

For simplicity, it is assumed that both sellers, A and B, know the discharging cost coefficient of the buyer, C.

Fig. 8 shows the normalized price of electricity (in Maui Island) as a function of time over the course of a single day, price updates resolve every fifteen minutes. The normalized average water consumption for the houses of aggregators *A* and *B* is shown in Fig. 9 for the same time intervals.

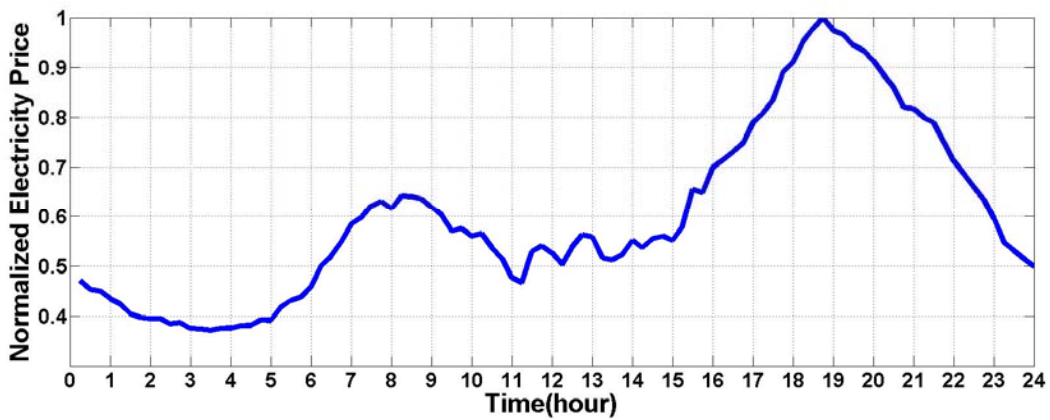

Fig. 8: Normalized electricity price in one day of case study (resolution: 15minutes)

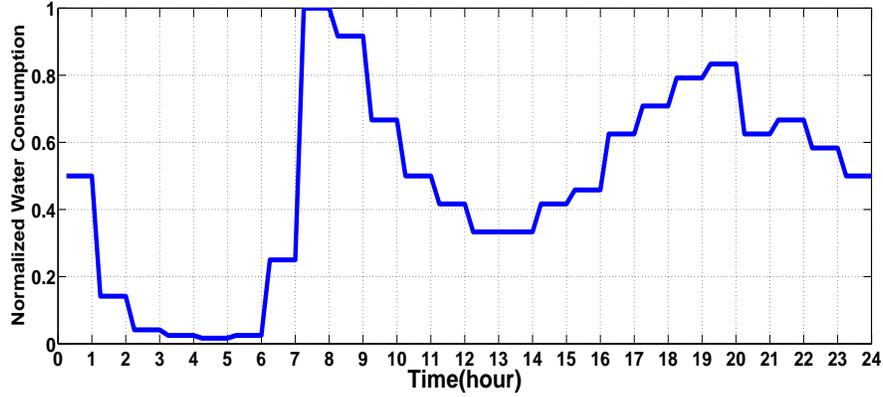

Fig. 9: Normalized average water consumption for houses of seller aggregators of case study (resolution: 15minutes)

Results for non-cooperative and Stackelberg games, with and without DR scheduling are presented in the following sections.

### 4.1. Results of non-cooperative game without DR scheduling

For our purposes, we take prices over half of the normalized price of electricity to be "expensive". From the normalized price data (Fig. 8), $P(elec_{exp}) = 0.7083, P(elec_{ch}) = 0.2917$. The status of the WHs (on/off) may be determined from the water consumption data (Fig. 9). In order to keep water temperatures in the range of $110$ to $130°F$, a typical WH runs (consumes power) for 18 of the 96 fifteen minute intervals of the day. Fig. 10 is the output of DP scheduler and shows the status of the seller's WHs in this game (non-cooperative without DR scheduling).

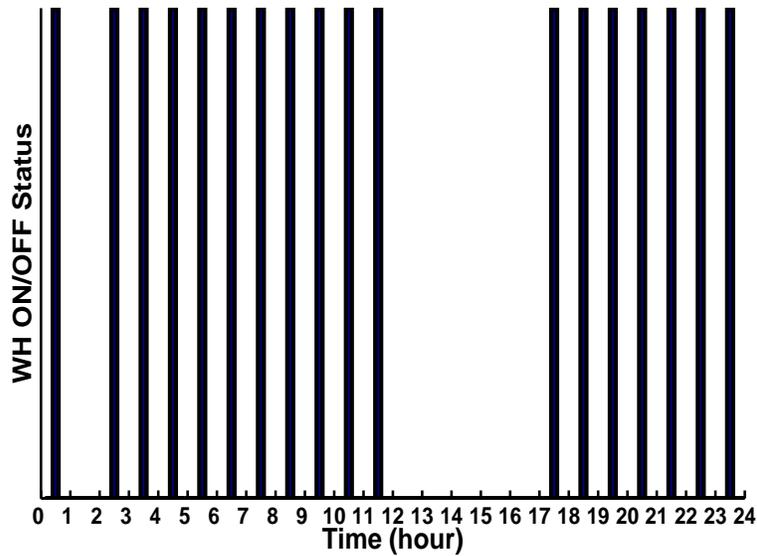

Fig. 10: On/Off status of WHs in non-cooperative game without DR scheduling for the seller aggregators

The probabilities $P(WH_{on}) = 0.1875$ and $P(WH_{off}) = 0.8125$ follow immediately from the known ratio of on to off intervals. Using the equations from section 3.2.1.1, the probabilities of the four scenarios are:
$P(\sigma_1) = 0.5755, P(\sigma_2) = 0.1328, P(\sigma_3) = 0.2370, P(\sigma_4) = 0.0547$.
Probability distributions for both DRAs *A* and *B* in the above mentioned scenarios are:

$\Psi_A^1 = [0.16, 0.84], \Psi_B^1 = [0.21, 0.79], \Psi_A^2 = [0.11, 0.89], \Psi_B^2 = [0.18, 0.82],$

$\Psi_A^3 = [0.75, 0.25], \Psi_B^3 = [0.67, 0.33], \Psi_A^4 = [0.69, 0.31], \Psi_B^4 = [0.60, 0.40].$

The expected probabilities (eq. 20) are:

$\pi_{11} = 0.0193$, $\pi_{12} = 0.0727$, $\pi_{21} = 0.1015$, $\pi_{22} = 0.3819$.

Conditional probability vectors for DRAs A and B are calculated using eq. 15 and 17:

$\eta_A^1 = [0.21, 0.79], \eta_B^1 = [0.16, 0.84], \eta_A^2 = [0.21, 0.79], \eta_B^2 = [0.16, 0.84]$.

We assume that each DRA choses one of the following strategies: biding 80% of marginal cost $(\varepsilon_{S1} = 1.6)$, biding marginal cost $(\varepsilon_{S2} = 2)$, and biding 120% of marginal cost $(\varepsilon_{S3} = 2.4)$. Therefore, the strategies are:

$s_A^1 = [0.00064, 0.00080, 0.00096]$

$s_A^2 = [0.00078, 0.00097, 0.00117]$

$s_B^1 = [0.00073, 0.00091, 0.00109]$

$s_B^2 = [0.00081, 0.00102, 0.00122]$

Finally, from the equations in step 5 of section 3.2.1.1, the expected pay off matrixes are:

$$EP_A^1 = \begin{bmatrix} 7.95 & 10.88 & 13.41 & 8.69 & 11.62 & 14.16 & 9.35 & 12.28 & 14.81 \\ 17.34 & 22.36 & 26.7 & 18.61 & 23.63 & 27.96 & 19.74 & 24.75 & 29.09 \\ 20.25 & 26.5 & 32 & 21.82 & 28.07 & 33.56 & 23.23 & 29.48 & 34.97 \end{bmatrix}$$

$$EP_A^2 = \begin{bmatrix} 4.12 & 6.71 & 9.06 & 4.75 & 7.34 & 9.69 & 5.34 & 7.93 & 10.28 \\ 11.62 & 16.09 & 20.12 & 12.72 & 17.19 & 21.22 & 13.74 & 18.21 & 22.24 \\ 13.13 & 18.57 & 23.55 & 14.45 & 19.89 & 24.87 & 15.7 & 21.14 & 26.12 \end{bmatrix}$$

$$EP_B^1 = \begin{bmatrix} 13.9 & 18.35 & 22.32 & 14.78 & 19.23 & 23.2 & 15.55 & 20 & 23.97 \\ 21.23 & 27.71 & 33.57 & 22.53 & 29.01 & 34.86 & 23.68 & 30.16 & 36.02 \\ 23.32 & 30.82 & 37.7 & 24.84 & 32.34 & 39.22 & 26.22 & 33.71 & 40.6 \end{bmatrix}$$

$$EP_B^2 = \begin{bmatrix} 53.57 & 89.85 & 117.01 & 59.86 & 96.14 & 123.3 & 64.56 & 100.84 & 128 \\ 25.18 & 63.56 & 92.7 & 31.92 & 70.3 & 99.44 & 37.03 & 75.41 & 104.55 \\ 0.61 & 40.01 & 70.22 & 7.59 & 46.99 & 77.2 & 12.94 & 52.34 & 82.56 \end{bmatrix}$$

The DRAs chose their bidding strategies by finding the Nash equilibrium. By inspecting the expected pay off matrices we find that for DRA A, $EP_A^1$ and $EP_A^2$, we see that the bottom row of each matrix, $s_A^1(3)$ and $s_A^2(3)$ - i.e. bid above marginal cost if A is of type 1 or type 2- results in the greatest expected payoff regardless of competitors strategies; we say that this strategy dominates the the others. DRA B will also conclude that a rational player A will choose strategy 3 (bid above marginal cost) regardless of its type. DRA B need only consider the best response to this strategy. This is done by considering the 9th column of $EP_B^1$ and $EP_B^2$. Which represents the payoffs of for B's possible strategies when competing against a player A who bids above marginal cost. Examining the last column of $EP_B^1$ and $EP_B^2$ tells us that: if B is type 1, it will receive the greatest payoff by bidding above marginal cost; and if B is type 2 it will receive the greatest payoff by bidding below marginal cost. Player A knows that B will play this way, but has no incentive to change its strategy. This strategy is represented by column 7th of $EP_A^1$ and $EP_A^2$ (column $\kappa_{31}$ of $EP_A$ in step 5 of section 3.2.1.1). The pair of strategies "9th column of $EP_B$" and "7th column of $EP_A$" is the Nash equilibrium of the game. The Nash equilibrium is a "consistent" prediction of how the game will be played by rational players. All participants predict that a particular Nash equilibrium will occur and there is no incentive to play differently. The strategy pairs in Nash equilibrium are the participants' maximum strategies which maximize participants' conditional payoffs. That is, a participant could obtain at least the payoff at the equilibrium point (or it may obtain more depending on his opponent's strategy).

In this section, scheduling of WHs are based on only one objective function: social welfare. The situation of each WH is based on water consumption only (Fig. 9) and is not effected by the electricity price (Fig. 8).

### 4.2. Results of non-cooperative game with price-sensitive WH scheduling

In this case, DP scheduler has been used to schedule the WHs of aggregator *A*, based on the electricity price (Fig.8) with the joint objectives of social welfare (keeping water temperature in a given range $[110,130]°F$ using Fig. 9) and DR scheduling using DP. Similar to section 4.1, $P(elec_{exp}) = 0.7083, P(elec_{ch}) = 0.2917$. Fig. 11 depicts scheduling of the WHs of aggregator *A* using DP (with inputs of data shown in Fig. 8 and 9). The highlighted areas show the times when the electricity is expensive (prices higher than half of maximum price).

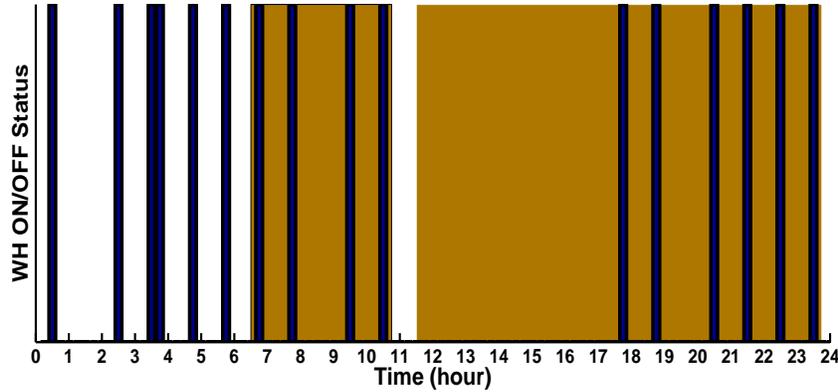

Fig. 11: On/Off status of WHs in non-cooperative game with DR scheduling for the aggregator *A*

In this case, based on the Fig. 11 and equations of section 3.2.1.2:
$P(\sigma_1) = P(elec_{exp}) * P(WH_{off} / elec_{exp}) = 0.7083 * 58/68 = 0.6041$
$P(\sigma_2) = P(elec_{exp}) * P(WH_{on} / elec_{exp}) = 0.7083 * 10/68 = 0.1042$
$P(\sigma_3) = P(elec_{ch}) * P(WH_{off} / elec_{ch}) = 0.2917 * 22/28 = 0.2292$
$P(\sigma_4) = P(elec_{ch}) * P(WH_{on} / elec_{ch}) = 0.2917 * 6/28 = 0.0625$

In aggregator *A*, Comparison of Fig. 10 with Fig. 11 shows DR scheduling can curtail the load of WHs as much as 11.2%. Therefore the new demand of the aggregator *A* is: $P_A^{demand} = 0.888 * 200 * 4.5 * 10/6 + P_A^{gen} = 1992 kW$. It is assumed here that aggregators *B* and *C* do not schedule their WHs. Therefore, the demands of these aggregators are the same as in section 4.1: $P_B^{demand} = 2394 kW, P_C^{demand} = 2478 kW$. Because of the load curtailment, aggregator *A* has more power stored in its batteries to sell to the aggregator *C*. With the same probability distributions and same strategies as in section 4.1 and following the same game process, the expected payoff matrices are:

$$EP_A^1 = \begin{bmatrix} 9.6 & 12.68 & 15.32 & 10.38 & 13.47 & 16.1 & 11.07 & 14.16 & 16.79 \\ 19.11 & 24.26 & 28.69 & 20.41 & 25.57 & 29.99 & 21.57 & 26.72 & 31.15 \\ 22.23 & 28.63 & 34.21 & 23.84 & 30.24 & 35.82 & 25.28 & 31.68 & 37.26 \end{bmatrix}$$

$$EP_A^2 = \begin{bmatrix} 5.7 & 8.49 & 10.99 & 6.39 & 9.18 & 11.68 & 7.02 & 9.82 & 12.31 \\ 13.35 & 18.02 & 22.19 & 14.5 & 19.17 & 23.34 & 15.56 & 20.23 & 24.4 \\ 15.09 & 20.75 & 25.87 & 16.48 & 22.14 & 27.26 & 17.77 & 23.43 & 28.56 \end{bmatrix}$$

$$EP_B^1 = \begin{bmatrix} 13.14 & 17.4 & 21.21 & 13.98 & 18.24 & 22.06 & 14.72 & 18.99 & 22.8 \\ 20.2 & 26.42 & 32.06 & 21.45 & 27.67 & 33.3 & 22.55 & 28.78 & 34.41 \\ 22.19 & 29.39 & 36.02 & 23.65 & 30.86 & 37.49 & 24.98 & 32.18 & 38.81 \end{bmatrix}$$

$$EP_B^2 = \begin{bmatrix} 45.74 & 81.66 & 108.55 & 51.96 & 87.88 & 114.77 & 56.62 & 92.53 & 119.42 \\ 17.57 & 55.54 & 84.37 & 24.23 & 62.2 & 91.03 & 29.29 & 67.25 & 96.09 \\ -6.73 & 32.23 & 62.11 & 0.16 & 39.12 & 69 & 5.45 & 44.41 & 74.29 \end{bmatrix}$$

Following the same reasoning used in section 4.1 for finding the Nash equilibrium, aggregator A obtains a higher payoff by bidding above its marginal cost. By inspecting the last column of $EP_B^1$ and $EP_B^2$, we learn that a rational participant B would bid $s_B^1(3)$ (above its marginal cost) when it is type 1 and $s_B^2(1)$ (below marginal cost) when it is type 2. This strategy is represented by column 7th of $EP_A^1$ and $EP_A^2$. The pair of strategies "9th column of $EP_B$" and "7th column of $EP_A$" is the Nash equilibrium of the game. The Nash equilibrium strategies are for A to bid higher than marginal cost regardless of its type and for B to bid above, or below marginal cost depending if it is type 1 or type 2, respectively.

### 4.3. Stackelberg game without DR scheduling

In non-cooperative game, there was no limitation power transactions or transaction prices for selling the stored energy between the aggregators. The transaction price and transaction power are obtained from Eq. 21, but in real markets there will exist a controller who limits the transaction power and transaction price. This independent controller is empowered by the utility and communicates with the utility to control the market. In this case study, for the non-cooperative games (section 4.1 and 4.2) the transaction prices (eq. 21) varied between $0.3342 and $0.4903. It is assumed now that the controller limits the selling price of the stored power to be less than $0.4 for each time interval. In the case of Stackelberg game without DR scheduling, the probabilities and strategies are the same as section 4.1, and the final expected pay off matrixes are:

$$EP_A^1 = \begin{bmatrix} 7.95 & 10.88 & 13.41 & 8.69 & 11.62 & 14.16 & 9.35 & 12.28 & 14.81 \\ 17.34 & 20.35 & 19.92 & 18.61 & 21.62 & 21.19 & 18.8 & 21.81 & 21.38 \\ 19.42 & 20.51 & 20.87 & 20.19 & 21.27 & 21.64 & 20.37 & 21.45 & 21.81 \end{bmatrix}$$

$$EP_A^2 = \begin{bmatrix} 4.12 & 5.03 & 2.67 & 4.75 & 5.66 & 3.29 & 4.5 & 5.41 & 3.04 \\ 10.36 & 9.67 & 8.56 & 10.55 & 9.86 & 8.75 & 10.34 & 9.65 & 8.53 \\ 9.97 & 10.39 & 10.2 & 10.2 & 10.62 & 10.42 & 10.23 & 10.65 & 10.45 \end{bmatrix}$$

$$EP_B^1 = \begin{bmatrix} 13.9 & 18.35 & 20.61 & 14.78 & 19.23 & 21.49 & 15.55 & 20 & 22.26 \\ 21.23 & 25.67 & 25.11 & 22.53 & 26.97 & 26.41 & 22.69 & 27.13 & 26.57 \\ 23.32 & 25.7 & 26.12 & 24.22 & 26.6 & 27.02 & 24.36 & 26.74 & 27.16 \end{bmatrix}$$

$$EP_B^2 = \begin{bmatrix} 53.57 & 87.11 & 107.5 & 59.86 & 93.4 & 113.79 & 63.85 & 97.38 & 117.78 \\ 23.5 & 54.03 & 75.74 & 29.05 & 59.58 & 81.29 & 32.81 & 63.33 & 85.05 \\ -4.67 & 26.81 & 49.41 & 0.8 & 32.28 & 54.88 & 4.72 & 36.2 & 58.8 \end{bmatrix}$$

Because of limitations for transaction power price, the expected payoff values have been decreased in this game in comparison with non-cooperative, unscheduled game (section 4.1). The Nash equilibrium strategies are still the same, by the same reasoning: for A to bid higher than marginal cost regardless of its type and for B to bid above, or below marginal cost depending if it is type 1 or type 2, respectively. The pair of strategies "9th column of $EP_B$" and "7th column of $EP_A$" is the Nash equilibrium of this game.

### 4.4. Stackelberg game with price-sensitive DR scheduling

In this case, aggregator A schedules its WHs based on the electricity price (Fig. 8) and water consumption (Fig. 9) and limitations on the market due to the constraints imposed by the controller must be considered This situation is the closest to reality in comparison with the other situations considered. The objective functions are both social welfare (keeping the water temperature of the houses in range of $[110,130]° F$ ) and DR scheduling (based on electricity price and water consumption). Also, the real market policies (with real limitations) have been applied on this situation. In the game, the probabilities and strategies are the same as section 4.2 and the final expected pay off matrixes are:

$$EP_A^1 = \begin{bmatrix} 9.60 & 12.68 & 15.32 & 10.38 & 13.47 & 16.10 & 11.07 & 14.16 & 16.79 \\ 19.11 & 23.32 & 23.01 & 20.41 & 24.63 & 24.32 & 20.92 & 25.13 & 24.82 \\ 22.13 & 23.32 & 23.78 & 23.13 & 24.32 & 24.78 & 23.33 & 24.52 & 24.98 \end{bmatrix}$$

$$EP_A^2 = \begin{bmatrix} 5.70 & 7.90 & 5.72 & 6.39 & 8.58 & 6.40 & 6.47 & 8.67 & 6.49 \\ 12.75 & 12.22 & 11.25 & 13.17 & 12.64 & 11.67 & 12.99 & 12.46 & 11.49 \\ 12.34 & 12.88 & 12.79 & 12.59 & 13.13 & 13.05 & 12.65 & 13.19 & 13.11 \end{bmatrix}$$

$$EP_B^1 = \begin{bmatrix} 13.14 & 17.40 & 21.01 & 13.98 & 18.24 & 21.86 & 14.72 & 18.99 & 22.60 \\ 20.20 & 25.53 & 25.11 & 21.45 & 26.77 & 26.36 & 21.86 & 27.18 & 26.77 \\ 22.19 & 25.36 & 25.88 & 23.25 & 26.42 & 26.94 & 23.41 & 26.58 & 27.10 \end{bmatrix}$$

$$EP_B^2 = \begin{bmatrix} 45.74 & 80.47 & 100.96 & 51.96 & 86.69 & 107.18 & 56.26 & 90.98 & 111.47 \\ 17.01 & 47.56 & 69.36 & 22.78 & 53.33 & 75.13 & 26.55 & 57.100 & 78.89 \\ -10.75 & 20.73 & 43.39 & -5.28 & 26.20 & 48.86 & -1.36 & 30.12 & 52.78 \end{bmatrix}$$

The Nash equilibrium strategy is for A to bid higher than marginal cost regardless of its type and for B to bid above, or below marginal cost depending if it is type 1 or type 2, respectively. Here, the pair of strategies "9th column of $EP_B$" and "7th column of $EP_A$" is the Nash equilibrium of this game.

Because of the presence of a controller in the Stackelberg games enforcing more market restrictions in the game, payoffs of aggregator A and B have been decreased in section 4.3 in comparison with 4.1 and also in section 4.4 in comparison with 4.2.

## 5. Conclusions

In this work we applied principles of elementary game theory to model competition between DRAs for selling power stored in electrochemical storage cells, in a power market with other DRAs as buyers. The situation presents itself to competitors as an incomplete information game as any DRA is unable to determine parameters associated with cost of operating other player's equipment. In order utilize the better developed theory of c-games with imperfect information we approximated local power demands using known water usage data by assuming that it was the major component of the local loads and using public statistical data and a Bayesian approach to derive probability distributions for DRA types. We then examined four types of c-game to consider the effects of pure competition (non-cooperative without WH scheduling) and regulated competition (Stackelberg without WH scheduling). We also considered both types of game with the addition of price-sensitive WH scheduling to examine the effect of adding DR to the games. The non-cooperative games are quite idealized and would not be practical to implement for various reasons explained above, but the Stackelberg games are relatively realistic and provide a possible beneficial alternative to currently popular market structures.

Application of our model to data compiled from the island of Maui served as a case study to analyze the results of our model under likely conditions. The results showed that bidding strategies of the DRAs was dependent only on parameters associated with hardware and not conditions of the games considered. Addition of price-sensitive DR WH scheduling increased payoffs to those DRAs which implemented it compared with those which did not. Payoffs were decreased in the regulated (Stackelberg) games compared with non-cooperative purer competition.

DR WH scheduling conserves energy and the resources spent acquiring it. Competition between DRAs for selling power independently of the generating company provides an alternative to the near-monopoly market structure currently prominent. Competition for selling power in a market decreases the market price of electricity and allows for smaller independent energy produces to enter the market thereby lowering the demands placed on the generating company. In our treatment, power stored in batteries is bought from the utility, but if the power was produced locally by renewable energy technologies the treatment would be analogous. Increasing penetration of renewable energy generating technologies such as photovoltaic systems will have a great effect on gaining independence from our combustible fossil-fuel economy and reducing pollution. We believe that proliferation and optimization of DR smart grid technologies will play a vital role in building a sustainable power economy.

The University of Hawaii's Renewable Energy Design Lab (REDlab) is currently developing hardware systems capable of implementing the presented model. Further work will be forthcoming exploring the implications of our game-theoretic competition model as an alternative market structure in the state of Hawaii where we currently face difficulties with integration of larger numbers of PV systems into the legacy grid. Much more work will be required if we are to reach the states 100% renewable energy goals.

**Acknowledgements**

This project is sponsored by the U.S. National Science Foundation under award number: 1310709. Thanks to for John Branigan's helping to edit this paper.